\documentclass[aps,prl,twocolumn,groupedaddress,showpacs]{revtex4}

\usepackage{graphicx} % uncomment this line to include the graphicx package
\usepackage{amssymb} % for the square root function

\usepackage{amsmath} %follow EK
\usepackage{dcolumn} %follow EK
\usepackage{bm}      %follow EK

\begin{document}

%\preprint{}

\title{Search for Superfluidity in Solid Hydrogen}


\author{A. C. Clark}
    \email{bg20339@hotmail.com}
\author{X. Lin}
\author{M. H. W. Chan}
%\thanks{}
\affiliation{Department of Physics, The Pennsylvania State
University, University Park, Pennsylvania 16802}


\date{\today}

\begin{abstract}

A torsional oscillator study of solid para-hydrogen has been carried
out down to 20 mK in search for evidence of superfluidity. This work
was inspired by the observation of the supersolid phase in solid
$^4$He. We found evidence of a possible phase transition, marked by
an abrupt increase in the resonant period of oscillation and onset
of extremely long relaxation times as the temperature was raised
above 60 mK. The change in the period for para-hydrogen, in contrast
to solid $^4$He, is not related to superfluidity as it fails a
crucial test showing that it is not a consequence of irrotational
flow. The long relaxation times observed suggest the effect is
related to the motion of residual ortho-hydrogen molecules in the
solid.

\end{abstract}

%\keywords{}

\maketitle

A supersolid is a solid that possesses an element of superfluidity.
The existence of supersolid $^4$He was first suggested 67 years ago
\cite{wolfke}, while possible underlying mechanisms were discussed
30 years later \cite{theory}. Experimental evidence of supersolid
$^4$He was reported by Kim and Chan (KC) only two years ago in a
series of torsional oscillator experiments on bulk solid helium
\cite{KC1,KCprl} and solid helium confined in porous media
\cite{KC2,PG}. KC demonstrated that the torsional oscillator is an
ideal technique for the detection of solid $^4$He acquiring
non-classical rotatonal inertia (\textit{NCRI}) \cite{tony}.
\textit{NCRI} is indicated by a drop in the resonant period of the
oscillator below the transition temperature. The decrement in the
period is proportional to the supersolid fraction in the limit of
low oscillation speed \cite{KC1}. These observations have since been
replicated in three other laboratories \cite{APS}. This prompted us
to address the question of whether or not the supersolid phase is
present in any system other than $^4$He.

Helium is the most quantum mechanical solid of all the elements.
This is due to the combination of large zero-point energy of
individual atoms and weak attractive interactions among them. A
measure of the quantum nature of a substance in the condensed
phases, i.e. liquid or solid, is given by the de Boer parameter,
$\Lambda$ = \textit{h}/$\sigma\sqrt{m\epsilon}$, where $\sigma$ and
$\epsilon$ are fitting parameters for the inter-particle potential
energy \cite{deboer}. Only helium and hydrogen isotopes have
$\Lambda$ greater than unity ($\Lambda_{3He}$ = 3.0, $\Lambda_{4He}$
= 2.6, $\Lambda_{H2}$ = 1.7, $\Lambda_{HD}$ = 1.4, $\Lambda_{D2}$ =
1.2). Therefore H$_2$, after $^4$He, is the most likely system to
exhibit superfluidity. Indeed the zero-point motion of solid H$_2$
at 5.4 K under zero pressure is 18\% of the nearest neighbor
distance \cite{neutron}, while in low density solid $^4$He it is
26\% \cite{xray}. Both of these values are significantly higher than
the Lindemann (melting) criterion of approximately 10\% for
classical solids. The fact that the supersolid phase in $^4$He
persists up to at least 135 bar \cite{KCprl}, where it is expected
to be less quantum mechanical than at lower pressures, was further
incentive to investigate solid H$_2$.

We have carried out a series of torsional oscillator measurements on
solid hydrogen. Our experimental configuration is similar to that
used by KC \cite{KC1}. A sample cell is suspended from the cooling
stage of a $^3$He-$^4$He dilution refrigerator by an annealed
beryllium-copper torsion rod. The rod provides a restoring force to
keep the cell in torsional motion and doubles as a filling line to
the cell. The H$_2$ space in the cell is comprised of an open
annulus and a central filling line, which are connected by three
channels that span the diameter of the cell. The annulus has an
average radius, width, and height of 6.5 mm, 2.3 mm, and 4.8 mm,
respectively. The radial channels and filling line have a diameter
of 1 mm, a combined volume that constitutes 7\% of the total cell
volume, and contribute about 2\% of the rotational inertia of all
the H$_2$ in the cell. Electrodes on the exterior of the torsion
cell are used to capacitively drive and detect oscillatory motion.
The resonant period ($\tau_0$) of the oscillator, is determined by
the moment of inertia (\textit{I}) of the torsion cell and the
torsion spring constant ($\kappa$) of the rod, such that $\tau_0$ =
2$\pi\sqrt {I/\kappa}$. The mechanical quality factor (\textit{Q})
of the oscillator at low temperature is 1.1 x 10$^6$, which allows
for the detection of changes in $\tau_0$ of one part in 10$^{7}$.

Solid H$_2$ samples were grown at saturated vapor pressure from high
purity H$_2$ gas containing very low levels of isotopic impurities:
less than 10 ppm of both hydrogen deuteride (HD) and deuterium
(D$_2$). However, for pure H$_2$ crystals there exist impurities of
a second kind. The two lowest rotational energy states of each
molecule are the para- (p-H$_2$) and ortho- (o-H$_2$) states, having
angular momentum \textit{J} = 0 and \textit{J} = 1, respectively.
The equilibrium concentrations of p-H$_2$ (1 - \textit{x}) and
o-H$_2$ (\textit{x}) at 300 K are respectively 0.25 and 0.75. At
temperatures (\textit{T}) below 4 K the equilibrium value of
\textit{x} is approximately zero and the solid is essentially pure
p-H$_2$. However, the conversion rate from o-H$_2$ to p-H$_2$ in the
solid is very slow for \textit{T} $>$ 500 mK (eg. about three weeks
of conversion are required for \textit{x} to drop from 0.01 to
0.009). This rate can be increased substantially by keeping H$_2$ in
the liquid phase (with \textit{T} $>$ 13.8 K, the triple point
temperature) in the presence of a high surface area, magnetic
material. In this way we attain \textit{x} $<$ 0.01. This process is
irreversible so that the average \textit{x} of each sample can only
decrease with time. Concentrations are determined by keeping the
mixing chamber at 20 mK and measuring the temperature difference
between the torsion cell and mixing chamber. Based on the
temperature difference and the thermal conductance of the Be-Cu
torsion rod, we can infer the amount of o-p conversion heat being
released and hence \textit{x} within the H$_2$ sample. The smallest
measurable temperature gradient corresponds to \textit{x} = 0.01.
All H$_2$ samples discussed in this work have \textit{x} $<$ 0.01
(i.e. no observed gradient). To limit the uncertainty in the
absolute value of \textit{x}, we measured the thermal conductivity
of our Be-Cu torsion rod and found good agreement with extrapolated
values from high temperature (\textit{T} $>$ 1 K) data previously
reported \cite{becu}.

\begin{figure}[b]
\includegraphics[width=0.9\columnwidth]{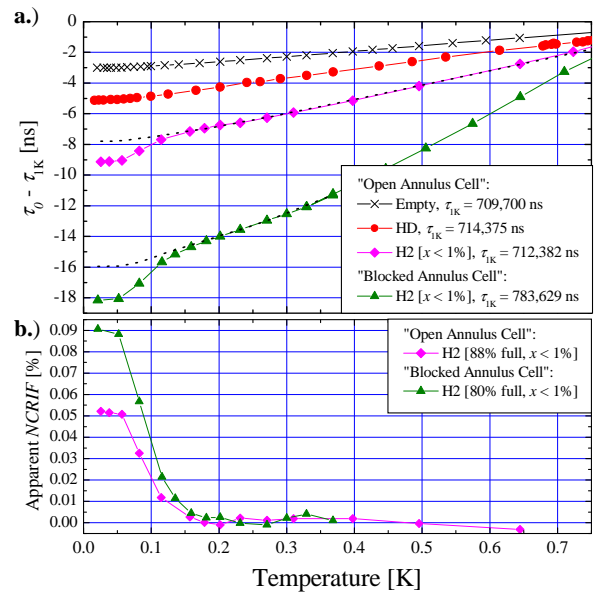}
\caption{\label{fig:one}\textbf{a.)} \textit{T}-dependence of the
resonant period for both the ``open'' and ``blocked'' cells. All of
the data shown were obtained at maximum rim velocities less than 40
$\mu$m/s. For the open annulus, data is plotted for the empty cell,
cell containing HD, and cell containing H$_2$. H$_2$ data taken
using the blocked cell exhibits the same qualitative features,
including long relaxation times. \textbf{b.)} \textit{T}-dependence
of the absolute period change from the high-\textit{T} background,
normalized by the total H$_2$ mass loading (``\textit{NCRIF},'' the
apparent \textit{NCRI} fraction). The \textit{T}-dependence is
similar to that observed in solid $^4$He. However, the same effect
is seen in the blocked cell.}
\end{figure}

In Fig.~\ref{fig:one}a we have plotted the resonant period as a
function of temperature for several samples. The empty cell period
at 1 K ($\tau_{1K}$) is 709,700 ns, which decreases by only 3 ns, or
five parts in 10$^{6}$, as the cell is cooled to 20 mK. This change
is determined by the \textit{T}-dependence, if any, of \textit{I}
and $\kappa$. Based on earlier work \cite{TO}, the small change in
$\tau_0$ depicted in Fig.~\ref{fig:one}a is due to small changes in
\textit{I} and/or $\kappa$ rather than large, nearly compensating
changes of each. For the H$_2$ sample in Fig.~\ref{fig:one}, the
period increases from the empty cell value by 2682 ns upon
solidification, indicating that the cell is 88\% full. The mass
loading due to the heavier and denser isotope, HD, is 4675 ns,
corresponding to 91\% filling. Each data set has been shifted
vertically for easy comparison. The corresponding values of
$\tau_{1K}$ for each are noted in the figure. There is a slight
difference in the slope of each curve shown. This modification of
the ``background'' when the sample is introduced into the cell is
always observed in torsional oscillator studies of liquid helium
films \cite{TO}, as well as in the solid helium work \cite{KC1,KC2}.

The most striking feature in the data of Fig.~\ref{fig:one}a is the
clear difference between H$_2$ and HD. The abrupt drop in $\tau_0$
below 180 mK for the cell containing H$_2$ is absent for HD. This
discrepancy is similar to that found between $^4$He and $^3$He, and
indicates that the drop in the period could be a signature of
\textit{NCRI}, as it is seen only in bosonic solids. The net change
in $\tau_0$ is obtained by subtracting the measured period from
values extrapolated from high temperature, as indicated in the
figure. While the period drop is more than an order of magnitude
smaller than that found in solid $^4$He in a similar geometry
\cite{KC1}, its temperature dependence is similar
[Fig.~\ref{fig:one}b]. The change in $\tau_0$ first becomes
measurable below 180 mK, and increases first gradually with
decreasing \textit{T} and then much more rapidly prior to saturation
near 60 mK. This similarity in the \textit{T}-dependence also
suggests that the period drop in H$_2$ could be a signature of
\textit{NCRI}.

However, there are four puzzling features in the H$_2$ data. First,
for solid $^4$He the drop in the period is accompanied by a minimum
in the amplitude of oscillation at the temperature where $\tau_0$
changes most rapidly \cite{KC1}. There is no such minimum in
oscillation amplitude near the transition region in the H$_2$
experiment. Second, for $^4$He the measured \textit{NCRIF} decreases
once the maximum oscillation speed of the sample exceeds 10 $\mu$m/s
\cite{KC1}. This speed has been interpreted as the critical velocity
of superflow. There is no evidence of a critical velocity in solid
H$_2$. We found the period drop to be independent of oscillation
speed up to 500 $\mu$m/s, the maximum within the linear response
range of the oscillator.

Third, the apparent supersolid onset temperature (180 mK) and
saturation temperature (60 mK) are remarkably close to the 230 mK
and 50 mK values found for solid $^4$He \cite{KC1}. The similarity
in these characteristic temperatures is inconsistent with the
substantial difference in $\Lambda_{4He}$ and $\Lambda_{H2}$. For
solid $^4$He in Vycor glass, very low $^3$He impurity levels can
significantly enhance the onset temperature \cite{KC2}. While the
\textit{T$_C$} for a sample containing 0.2 ppm of $^3$He is about
175 mK, a sample with 10 ppm has an onset temperature around 370 mK.
This effect has also been observed in bulk $^4$He \cite{He3}, thus a
small amount of HD impurities may alter \textit{T$_C$} for solid
H$_2$. However, we do not know the exact HD impurity level in our
solid H$_2$ samples. The reason is that HD is heavier than H$_2$ and
hence preferentially adsorbs onto surfaces, which are abundant in
the H$_2$ gas handling system leading into the torsion cell.
Specifically, the capillary system contains a chamber filled with a
high surface area, FeO(OH) compound. Thus, we expect that the actual
HD concentration in each sample is much less than the quoted value
of 10 ppm. Since we do not know with certainty the exact HD
concentration we cannot validate the 180 mK onset and 60 mK
saturation temperatures of \textit{NCRI}.

\begin{figure}[t]
\includegraphics[%
width=0.9\columnwidth ]{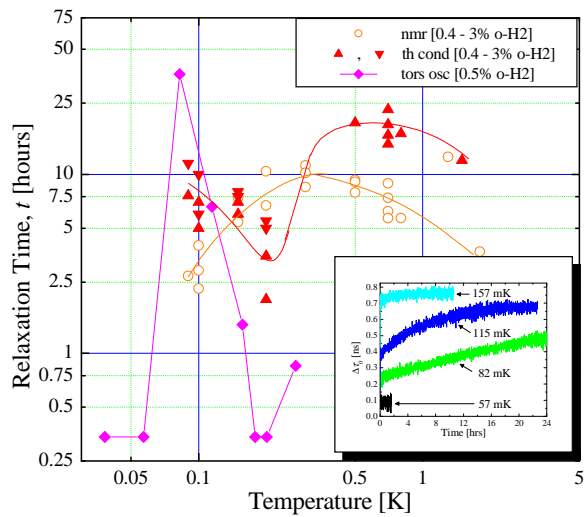}
\caption{\label{fig:two}\textit{T}-dependence of measured relaxation
times, including data from the literature \cite{K}. The sharp
increase in \textit{t} coincides with the observed period shift. At
\textit{T} $\sim$ 180 mK, a minimum in \textit{t} is observed for
both the torsional oscillator and thermal conductivity measurements.
Inset: Time dependence of $\tau_0$ at several different
temperatures. The curves have been shifted vertically for clarity.
The longest relaxation times are observed at \textit{T} $\sim$ 80
mK.}
\end{figure}

The final and most unusual feature of the H$_2$ data, compared to
$^4$He, is the exceedingly long relaxation time (\textit{t}) of the
period reading [Fig.~\ref{fig:two}]. It is expected that the
oscillator comes into thermal and mechanical equilibrium in the
following manner. The e$^{-1}$ time constant required for thermal
equilibration increases with decreasing \textit{T} and typically
reaches 15 minutes below 50 mK. Since the mechanical \textit{Q} is
on the order of 10$^6$, the necessary time to allow for mechanical
equilibrium is about 10 minutes. Therefore it is necessary to wait
at least 75 minutes after each temperature step before collecting
$\tau_0$ data. This protocol was used in the solid helium
experiments to ensure reproducible data upon warming and cooling.
However, in gathering data on H$_2$ we observe \textit{t}-values as
long as 36 hours just above 60 mK. There is no such evidence of
relaxation in the HD sample or for the empty cell, for which
$\tau_0$ equilibrates within 75 minutes at all \textit{T}. We will
return to the issue of \textit{t}-values below.

We have carried out a crucial control experiment to ascertain
whether the observed effect in solid H$_2$ is a signature of
superfluidity. We constructed another sample cell, identical to the
first except for the addition of a barrier at one point in the
annulus. Its purpose is to make the path for the irrotational
superflow much more tortuous, leading to a predictable decrement in
\textit{NCRI} \cite{flow1,flow2}, thus reducing the drop in the
period. This control experiment provides a convincing case for the
supersolid phase of $^4$He. It was found that when the annulus was
blocked, the drop in $\tau_0$ was reduced by a factor of 70 from the
open annulus value \cite{KC1}. This reduction is in agreement with
what is expected for irrotational superflow \cite{flow2}. The result
of this control experiment for solid H$_2$ is plotted alongside the
data of the ``open annulus cell'' in Fig.~\ref{fig:one}a. The period
drop below 180 mK is still present for the ``blocked annulus cell.''
In fact, we find the magnitude is larger in the latter. One possible
reason for this is that, while each sample has \textit{x} $<$ 0.01,
the exact ortho-concentrations of the two samples may differ.
Regardless, the result of the control experiment clearly indicates
that the H$_2$ result is completely different from that found in
solid $^4$He and is not related to superfluidity. We note that, as
in the open cell, long relaxation times are observed just above the
saturation temperature of 60 mK.

Long equilibration times are typical in studies of o-H$_2$ diffusion
in solid p-H$_2$ \cite{reviews}. In the dilute limit where
\textit{x} is less than several percent, isolated o-H$_2$ molecules
(singles) cluster together below 1 K even though thermally activated
diffusion is severely limited. The o-H$_2$ impurities effectively
propagate throughout the lattice via the transferral of \textit{J}
from molecule to molecule \cite{resOP}, rather than by particle
exchange. The ``equilibrium'' state of the solid thus involves a
temperature dependent distribution of singles, pairs (two
neighboring o-H$_2$ molecules) and larger clusters of o-H$_2$. At
several kelvin, all of the o-H$_2$ molecules are randomly
distributed as singles. As the temperature is lowered the
concentration of isolated singles decreases in favor of the
formation of pairs and larger clusters. Since o-H$_2$ and p-H$_2$
have different densities, in the limit of zero temperature the
system prefers to be in a macroscopically phase-separated state.

Clustering was first observed by monitoring the time evolution of
the nuclear magnetic resonance (nmr) absorption spectra for o-H$_2$
singles and pairs after rapid warming or cooling of solid H$_2$
crystals \cite{NMR1}. Representative growth (above 1 K) and decay
(below 1 K) rates of the singles spectrum \cite{K} are presented in
Fig.~\ref{fig:two}. The singles decay rate exhibits a non-monotonic
\textit{T}-dependence, decreasing dramatically between 300 mK and
100 mK. Well below 100 mK the decay of singles is on a time scale of
minutes rather than hours, rendering the measurement impossible. The
pair spectrum also shows an enhancement in its decay rate at low
\textit{T} but time constants are at least one order of magnitude
longer, eg. on the order of 10 hours at 25 mK \cite{NMR2}. Also
included in Fig.~\ref{fig:two} are relaxation times obtained from
simultaneous thermal conductivity measurements \cite{K}. The time
evolution of the thermal conductivity depends strongly on
\textit{T}. For \textit{T} $>$ 300 mK the time dependence is
described by a simple exponential. However, contrary to the singles
decay rate, a sum of two exponentials best fits the conductivity
data below 300 mK. This difference may be related to the presence of
o-H$_2$ clusters at lower temperature. Since it is difficult to
resolve the spectra of triples or larger clusters \cite{NMR3}, nmr
studies cannot quantify the degree of mobility of larger o-H$_2$
clusters that form at low \textit{T}. Their presence, however, has
been detected for solid H$_2$ crystals held at 25 mK for only 24
hours \cite{NMR4}.

Our torsional oscillator data are obtained using the following
protocol. Upon cooling H$_2$ samples to 20 mK, one to two weeks are
allowed for equilibration, during which time $\tau_0$ drops smoothly
until finally stabilizing to within 0.05 ns, our limiting
resolution. We note that the rate of irreversible o-p conversion is
enhanced for \textit{T} $<$ 200 mK since local \textit{x}-values are
higher in clustered regions \cite{NMR5}. However, upon stabilizing
at 20 mK, we observe no long term drift in the $\tau_0$-reading.
After complete equilibration the temperature is raised in successive
steps, for each of which $\tau_0$ is measured as a function of time
[Fig.~\ref{fig:two} inset]. We find for all samples that $\tau_0$
equilibrates quickly for any \textit{T} $<$ 60 mK. However, further
increase in \textit{T} leads to extremely slow relaxation. Near and
above 180 mK we find that \textit{t} shortens, returning to a value
of about 15 minutes, the usual e$^{-1}$ equilibration time. When the
temperature sweep is complete the system is returned to 20 mK and
allowed to re-equilibrate. Our measured \textit{t}-values are
presented in Fig.~\ref{fig:two}.

Associating the motion of o-H$_2$ impurities with the observed time
dependence of $\tau_0$ in the present work explains the absence of
this unusually long relaxation in HD, $^4$He, and $^3$He. In our
experiment we allow clustering to ensue for up to two weeks at 20
mK. Based on measured decay times in the literature, no isolated
o-H$_2$ singles or pairs remain in our samples at the time we begin
each temperature sweep. Thus, we are able to follow the
un-clustering process upon gradual warming of the torsion cell.

The moment of inertia and hence $\tau_0$ are extremely sensitive to
the radial density profile of the sample. Changes in the period
require the rearrangement of o-H$_2$ within the torsion cell in such
a way as to reduce or enhance the density at large radii. We think
the most likely explanation for the observed drop in $\tau_0$ below
180 mK is the formation of large clusters of nearly pure o-H$_2$ in
the center of the cell (fill line and radial channels), increasing
the purity of p-H$_2$ in the annulus. Since the density of the
o-H$_2$ is 1.7\% higher than that of p-H$_2$, the clustering of
o-H$_2$ toward the center has the net effect of reducing \textit{I}.
An extreme case is where all of the o-H$_2$ molecules cluster to the
center of the torsion cell (the filling line). In this situation,
the reduction of $\tau_0$ is 1.7 x 10$^{-4}$ of the total H$_2$ mass
loading. This is on the same order of our observations
[Fig.~\ref{fig:one}], i.e. 5.5 x 10$^{-4}$ and 9.0 x 10$^{-4}$ in
the open and blocked cells, respectively. It is known that solid
H$_2$ does not wet many metal substrates \cite{wet}. If the contact
angle of the sample to the cell walls is different for the mixed and
phase-separated phases (eg. p-H$_2$), \textit{I} will show a change
upon phase separation. This is one possible explanation for larger
than expected change in $\tau_0$ that we observe.

In conclusion, a resonant period shift in our torsional oscillator
containing solid H$_2$ is observed. While the phenomenon shares some
features with the supersolid transition in $^4$He, there are several
dissimilarities. Most notable is the presence of the period drop in
both the open and blocked cells, proving that the effect is not
related to superfluidity. We believe the abrupt rise in $\tau_0$
accompanied by a sudden increase in \textit{t} near 60 mK is
consistent with a transition from a phase-separated configuration to
one that is un-clustered at high temperature.

\begin{acknowledgments}
We thank E. Kim and W. N. Hardy for their advice. Special thanks to
H. Meyer for many enlightening discussions. Support comes from NSF
Grant DMR 0207071.
\end{acknowledgments}

%\bibliography{H2TObib}

\begin{thebibliography}{25}
\bibitem{wolfke} M. Wolfke, Ann. Acad. Sci. techn. Varsovie {\textbf 6}, 14
(1939).
\bibitem{theory} A. F. Andreev and I. M. Lifshitz, Sov. Phys. JETP \textbf{29}, 1107
(1969); G. V. Chester, Phys. Rev. A \textbf{2}, 256 (1970).
\bibitem{KC1} E. Kim and M. H. W. Chan, Science \textbf{305}, 1941 (2004).
\bibitem{KCprl} E. Kim and M. H. W. Chan, Phys. Rev. Lett. \textbf{97}, 115302 (2006).
\bibitem{KC2} E. Kim and M. H. W. Chan, Nature \textbf{427}, 225 (2004).
\bibitem{PG} E. Kim and M. H. W. Chan, J. Low Temp. Phys. \textbf{138}, 859 (2005).
\bibitem{tony} A. J. Leggett, Phys. Rev. Lett. \textbf{25}, 1543
(1970).
\bibitem{APS} A. S. C. Rittner and J. D. Reppy, cond-mat 0604528 (2006); M. Kondo \textit{et al.}, cond-mat 0607032 (2006);
A. Penzyev and M. Kubota, private communication.
\bibitem{deboer} J. de Boer, Physica \textbf{14}, 139 (1948).
\bibitem{neutron} M. Nielsen, Phys. Rev. B \textbf{7}, 1626 (1973).
\bibitem{xray} C. A. Burns and E. D. Isaacs, Phys. Rev. B \textbf{55}, 5767 (1997).
\bibitem{becu} V. Gr\"oger and F. Stangler, Z. Metallkd. \textbf{72}, 487 (1981).
\bibitem{TO} G. Agnolet, D. F. McQueeney, and J. D. Reppy, Phys. Rev. B \textbf{39}, 8934 (1989).
\bibitem{He3} E. Kim and M. H. W. Chan, private communication.
\bibitem{flow1} A. L. Fetter, J. Low Temp. Phys. \textbf{16}, 533 (1974).
\bibitem{flow2} E. J. Mueller, private communication (2005).
\bibitem{reviews} H. Meyer, Can. J. Phys. \textbf{65}, 1453 (1987); Low Temp. Phys. \textbf{24}, 381 (1998).
\bibitem{resOP} R. Oyarzun and J. Van Kranendonk, Phys. Rev. Lett.
\textbf{26}, 646 (1971).
\bibitem{NMR1} L. I. Amstutz, J. R. Thompson, and H. Meyer, Phys. Rev. Lett. \textbf{21}, 1175 (1968).
\bibitem{K} X. Li, D. Clarkson, and H. Meyer, J. Low Temp. Phys. \textbf{78}, 335 (1990).
\bibitem{NMR2} S. Washburn, R. Schweizer, and H. Meyer, J. Low Temp. Phys. \textbf{40}, 187 (1980).
\bibitem{NMR3} A. B. Harris \textit{et al.}, Phys. Rev. \textbf{175}, 603 (1968).
\bibitem{NMR4} R. Schweizer \textit{et al.}, J. Low Temp. Phys. \textbf{37}, 309 (1979).
\bibitem{NMR5} R. Schweizer \textit{et al.}, J. Low Temp. Phys. \textbf{37}, 289 (1979).
\bibitem{wet} M. Sohaili, J. Klier, and P. Leiderer, J. Phys.: Condens. Matter \textbf{17}, S415 (2005).
\end{thebibliography}

\end{document}